\documentclass[12pt]{article}
\usepackage[pctex32]{graphicx}
\linethickness{0.4pt}

\begin{document}
\vskip 2.0 true cm 
\begin{center}
{\LARGE \bf Long-range Effects on the Pyroelectric Coefficient and Dielectric Susceptibility of a Ferroelectric Bilayer}
\vskip 0.5     true cm
{\large Dong-Lai Yao,$^{b)}$ Yin-Zhong Wu,$^{b), c)}$ Wen Dong,$^{b)}$ and Zhen-Ya Li$^{a), b)}$}
\vskip 0.2 true cm

{\it $^{a)}$ CCAST(World Laboratory), P.O.Box 8730,Beijing 100080, People's
Republic of China}\\
{\it $^{b)}$ Department of Physics, Suzhou University, Suzhou 215006, China$^*$}\\
{\it $^{c)}$ Department of Physics, Changshu College, Changshu 215500, China}\\
\end{center}
\begin{abstract}
Long-range effects on the pyroelectric coefficient and susceptibility of a ferroelectric bilayer with a ferroelectric interfacial coupling are investigated by use of the transverse Ising model within the framework of mean-field theory. The effects of the interfacial coupling and the transverse field on the pyroelectric coefficient and susceptibility of the bilayer are investigated by taking into account the long-range interaction. It is found that the pyroelectric coefficient and susceptibility increase with the decrease of the magnitude of the long-range interaction and the interfacial coupling when the temperature is lower than the phase transition temperature. We also find that the strong long-range interaction, the large transverse field and weak interfacial coupling can lead to the disappearance of some of the peaks of the pyroelectric coefficient and susceptibility of the ferroelectric bilayer. The phase transition temperature increases with the increase of the strength of the long-range interaction, which is similar to the results obtained in ferroelectric multi-layers or superlattice.
\end{abstract}

$*$ Mailing Address in China\\
Electronic mail: dlyao$@$pub.sz.jsinfo.net or zyli$@$suda.edu.cn
\newpage
{\bf{I. INTRODUCTION}}\\

Ferroelectric, pyroelectric and piezoelectric properties of ferroelectric thin film have been extensively studied for their potential applications experimentally and theoretically[1-3]. Many experiments showed that the interfacial coupling may have nontrivial effects on the properties of ferroelectric thin films, such as [4-5], $Pb$ was used as an interfacial layer to improve the ferroelectric property of the lead zirconate titanate (PZT) thin film.[6] The properties of the ferroelectric superlattice depend sensitively on the interfacial coupling and the thicknesses of the components.[7] Ma et al.[8] considered a superlattice consisting of two kinds of ferroelectric materials within the framework of Landau phenomenological theory. They assumed that the interfacial coupling between two ferroelectric materials was ferroelectric or antiferroelectric, and investigated the effects of the interfacial coupling on the polarization and the phase transition temperature of the ferroelectric superlattice. Sepliarsky et al. described a atom-scale simulation about the ferroelectric film and showed that the long-range effect in the ferroelectric film is important.[9]\\

Recently, the study of the pyroelectric effects of ferroelectric materials was active in both experiments[10-13] and theory[14-17]. Xin[14] et al. studied the pyroelectric coefficient of ferroelectric superlattice with ferroelectric interfacial coupling by use of transverse Ising Model. The long-range interaction in the ferroelectric film is thought to be unneglectable[18]. As is known, the bilayer structure is one of the basic models in the ferroelectric film. While the pyroelectric properties and the dielectric susceptibility of a ferroelectric bilayer with long-range coupling interaction have not been investigated so far. In this paper, the long-range interaction is taken into account in studying the pyroelectric coefficient and the dielectric susceptibility of the bilayer. The effects of the long-range interaction, the interfacial coupling, the transverse field and the temperature on the pyroelectric coefficient and dielectric susceptibility of a bilayer are investigated in detail. We find the geometric size effect on the pyroelectric coefficient and dielectric susceptibility under the strong long-range interaction coupling. The size-dependence of the phase transition temperature of the bilayer is also examined.\\

{\bf{II. MODEL AND FORMULATION}}\\

We consider a bilayer structure composed of two different ferroelectric slabs(A and B) with different number of layers, each layer is defined on the x-y plane and pseudo spins site on a square lattice(see Fig. 1). The system is described by the Ising Hamiltonian in the presence of a transverse field:
\begin{equation}
H=-\sum_{<ij>}J_{ij}S_{i}^{z}S_{j}^{z}-\sum_{i}\Omega_{i}S_{i}^{x}-2\mu E\sum_{i}S_{i}^{z},
\end{equation}
where $\Omega_{i}$ is the transverse field at site $i$. $S_{i}^{z}$ and $S_{i}^{x}$ are components of spin-1/2 operator at site $i$, $\mu$ is the dipole moment on site $i$, and $E$ is the applied electric field. The long-range interaction parameter $J_{ij}$ and the transverse field $\Omega_{i}$ are described as 
\begin{equation}
J_{ij}=\left \{
\begin{array}{ll}
\frac{\textstyle {J_{a}}}{\displaystyle {r_{ij}^{\sigma}}}, &{\rm for }\ i, j\in slab A, \nonumber\\\nonumber
\frac{\textstyle {J_{ab}}}{\displaystyle {r_{ij}^{\sigma}}}, &{\rm for }\ i\in slab A, j\in slab B, \\\nonumber
\frac{\textstyle {J_{b}}}{\displaystyle {r_{ij}^{\sigma}}}, &{\rm for }\ i, j\in slab B, \nonumber
\end{array}
\right.
\end{equation}

\begin{equation}
\Omega_{i}=\left \{
\begin{array}{ll}
\Omega_{a}, &{\rm for }\ i\in slab A, \nonumber \\\nonumber
\Omega_{b}, &{\rm for }\ i\in slab B, \nonumber
\end{array}
\right.
\end{equation}
where  $J_{a}$, $J_{b}$ and $J_{ab}$ are the nearest neighbor coupling constants in slab $A$, slab $B$ and between slab $A$ and $B$ respectively. $r_{ij}$ is the distance between site $i$ and $j$, $\sigma$ is introduced to describe the magnitude of the long-range interaction. $\Omega_{a}$ and $\Omega_{b}$ are the transverse fields in Slab A and Slab B, respectively. If $\sigma\longrightarrow \infty$, all the interactions in the bilayer will recover the short-range interactions. The magnitude of the long-range interaction will increase as $\sigma$ decreases. $\sigma = 0$ is corresponding to the infinite-range coupling. But considering the Coulomb-interaction nature of ferroelectric materials and that the phase transition temperature will diverge at $\sigma = 0$[19]. Thus, in our work, we take $\sigma = 3.0, 6.0$, and $\infty$ to study the long-range interaction effect. It must be emphasized that the interactions of the pseudo spins in slab $A$, slab $B$ and the interfacial coupling considered here are all long-range.  
Considering the environment of the sites at the same layer is identical, we assume that the average value of the pseudo spins in the same layer have the same value. We use an improved mean field theory, which take the correlation of the pseudo spins within the range of 8-th neighbor pseudo spins into consideration. Within such approximation, the average pseudo spin along $z$ direction in the $i-th$ layer can be expressed as following:[20-21]
\begin{equation}
R_{i}=<S_{i}>=\frac{<H_{i}^{z}>}{2|H_{i}|}\tanh{\frac{|H_{i}|}{2k_{B}T}},
\end{equation}
where
\begin{equation}
<H_{i}^{z}>=\sum_{j}J_{ij}R_{j}+2\mu E ,
\end{equation}

\begin{equation}
|H_{i}|=\sqrt {\Omega_{i}^{2}+(<H_{i}^{z}>)^{2} },
\end{equation}
where $R_{i}$ stand for the average polarization of $i-th$ layer, which is equivalent to the average value of pseudo spins, and $j$ runs over all the sites within the two slabs. In order to make the computation practicable, the long-range interactions are cut off at the eighth-neighbor in our calculations, we will discuss the practicability of the cut-off approximation in Sec. III. When $i$ runs over all the layers in the structure, the above Eq. (4) forms a set of simultaneous nonlinear equations from which $R_{i}$ can be calculated numerically.\\

For instance, we take $L_{a}=3, L_{b}=3$. And the long-range interaction is cut off at the eighth-neighbor, which indicates that we take $r_{ij}$ as $1$, $\sqrt 2$, $\sqrt 3$, 2, $\sqrt 5$, $\sqrt 6$, $2\sqrt 2$, and $3$. In such approximation, if $|r_i-r_j|\leq3$, the simultaneous nonlinear equations will be:

\[
H_1^z  = \sum\limits_{\scriptstyle i \in 1 \hfill \atop 
  \scriptstyle j \in 1 \hfill} {\frac{{J_a }}{{r_{ij}^\sigma  }}R_1 }  + \sum\limits_{\scriptstyle i \in 1 \hfill \atop 
  \scriptstyle j \in 2 \hfill} {\frac{{J_a }}{{r_{ij}^\sigma  }}R_2 }  + \sum\limits_{\scriptstyle i \in 1 \hfill \atop 
  \scriptstyle j \in 3 \hfill} {\frac{{J_a }}{{r_{ij}^\sigma  }}R_3 }  + \sum\limits_{\scriptstyle i \in 1 \hfill \atop 
  \scriptstyle j \in 4 \hfill} {\frac{{J_{ab} }}{{r_{ij}^\sigma  }}R_4 }  + 2\mu E,
\]
\[
H_2^z  = \sum\limits_{\scriptstyle i \in 2 \hfill \atop 
  \scriptstyle j \in 1 \hfill} {\frac{{J_a }}{{r_{ij}^\sigma  }}R_1 }  + \sum\limits_{\scriptstyle i \in 2 \hfill \atop 
  \scriptstyle j \in 2 \hfill} {\frac{{J_a }}{{r_{ij}^\sigma  }}R_2 }  + \sum\limits_{\scriptstyle i \in 2 \hfill \atop 
  \scriptstyle j \in 3 \hfill} {\frac{{J_a }}{{r_{ij}^\sigma  }}R_3 }  + \sum\limits_{\scriptstyle i \in 2 \hfill \atop 
  \scriptstyle j \in 4 \hfill} {\frac{{J_{ab} }}{{r_{ij}^\sigma  }}R_4 }  + \sum\limits_{\scriptstyle i \in 2 \hfill \atop 
  \scriptstyle j \in 5 \hfill} {\frac{{J_{ab} }}{{r_{ij}^\sigma  }}R_5 }  + 2\mu E,
\]
\[
H_3^z  = \sum\limits_{\scriptstyle i \in 3 \hfill \atop 
  \scriptstyle j \in 1 \hfill} {\frac{{J_a }}{{r_{ij}^\sigma  }}R_1 }  + \sum\limits_{\scriptstyle i \in 3 \hfill \atop 
  \scriptstyle j \in 2 \hfill} {\frac{{J_a }}{{r_{ij}^\sigma  }}R_2 }  + \sum\limits_{\scriptstyle i \in 3 \hfill \atop 
  \scriptstyle j \in 3 \hfill} {\frac{{J_a }}{{r_{ij}^\sigma  }}R_3 }  + \sum\limits_{\scriptstyle i \in 3 \hfill \atop 
  \scriptstyle j \in 4 \hfill} {\frac{{J_{ab} }}{{r_{ij}^\sigma  }}R_4 }  + \sum\limits_{\scriptstyle i \in 3 \hfill \atop 
  \scriptstyle j \in 5 \hfill} {\frac{{J_{ab} }}{{r_{ij}^\sigma  }}R_5 }  + \sum\limits_{\scriptstyle i \in 3 \hfill \atop 
  \scriptstyle j \in 6 \hfill} {\frac{{J_{ab} }}{{r_{ij}^\sigma  }}R_6 }  + 2\mu E,
\]
\[
H_4^z  = \sum\limits_{\scriptstyle i \in 4 \hfill \atop 
  \scriptstyle j \in 1 \hfill} {\frac{{J_{ab} }}{{r_{ij}^\sigma  }}R_1 }  + \sum\limits_{\scriptstyle i \in 4 \hfill \atop 
  \scriptstyle j \in 2 \hfill} {\frac{{J_{ab} }}{{r_{ij}^\sigma  }}R_2 }  + \sum\limits_{\scriptstyle i \in 4 \hfill \atop 
  \scriptstyle j \in 3 \hfill} {\frac{{J_{ab} }}{{r_{ij}^\sigma  }}R_3 }  + \sum\limits_{\scriptstyle i \in 4 \hfill \atop 
  \scriptstyle j \in 4 \hfill} {\frac{{J_b }}{{r_{ij}^\sigma  }}R_4 }  + \sum\limits_{\scriptstyle i \in 4 \hfill \atop 
  \scriptstyle j \in 5 \hfill} {\frac{{J_b }}{{r_{ij}^\sigma  }}R_5 }  + \sum\limits_{\scriptstyle i \in 4 \hfill \atop 
  \scriptstyle j \in 6 \hfill} {\frac{{J_b }}{{r_{ij}^\sigma  }}R_6 }  + 2\mu E,
\]
\[
H_5^z  = \sum\limits_{\scriptstyle i \in 5 \hfill \atop 
  \scriptstyle j \in 2 \hfill} {\frac{{J_{ab} }}{{r_{ij}^\sigma  }}R_2 }  + \sum\limits_{\scriptstyle i \in 5 \hfill \atop 
  \scriptstyle j \in 3 \hfill} {\frac{{J_{ab} }}{{r_{ij}^\sigma  }}R_3 }  + \sum\limits_{\scriptstyle i \in 5 \hfill \atop 
  \scriptstyle j \in 4 \hfill} {\frac{{J_b }}{{r_{ij}^\sigma  }}R_4 }  + \sum\limits_{\scriptstyle i \in 5 \hfill \atop 
  \scriptstyle j \in 5 \hfill} {\frac{{J_b }}{{r_{ij}^\sigma  }}R_5 }  + \sum\limits_{\scriptstyle i \in 5 \hfill \atop 
  \scriptstyle j \in 6 \hfill} {\frac{{J_b }}{{r_{ij}^\sigma  }}R_6 }  + 2\mu E,
\]
\[
H_6^z  = \sum\limits_{\scriptstyle i \in 6 \hfill \atop 
  \scriptstyle j \in 3 \hfill} {\frac{{J_{ab} }}{{r_{ij}^\sigma  }}R_3 }  + \sum\limits_{\scriptstyle i \in 6 \hfill \atop 
  \scriptstyle j \in 4 \hfill} {\frac{{J_b }}{{r_{ij}^\sigma  }}R_4 }  + \sum\limits_{\scriptstyle i \in 6 \hfill \atop 
  \scriptstyle j \in 5 \hfill} {\frac{{J_b }}{{r_{ij}^\sigma  }}R_5 }  + \sum\limits_{\scriptstyle i \in 6 \hfill \atop 
  \scriptstyle j \in 6 \hfill} {\frac{{J_b }}{{r_{ij}^\sigma  }}R_6 }  + 2\mu E.
\]

From the above equations, we can see that the mean field (or local field) of pseudo-spin $S_{1}^z$ is related to the mean value of $R_1$, $R_2$, $R_3$ and $R_4$ (according to our cut-off approximation). Therefore, the correlation between the pseudo spins has been partially considered in our calculation.\\

The polarization of the $ith$ layer is 
\begin{equation}
P_{i}= 2n\mu R_{i},
\end{equation}
where $n$ is the number of pseudo spins in a unit volume. The average polarization of the bilayer is obtained:
\begin{equation}
P_{av}=\frac{1}{N}\sum_{i=1}^{N}P_{i},
\end{equation}
where $N$ is the total layers of the two slabs.\\

The pyroelectric coefficient of the bilayer structure is defined as:
\begin{equation}
p(T)=-\frac{\partial P_{av}(T)}{\partial T}=-\frac{1}{N}\sum_{i=1}^{N}\frac{\partial P_{i}(T)}{\partial T}=-\frac{1}{N}\sum_{i=1}^{N} 2n\mu\frac{\partial R_{i}(T)}{\partial T}.
\end{equation}

The dielectric susceptibility of the bilayer structure is defined as:
\begin{equation}
\chi (T)=\frac{\partial P_{av}(E,T)}{\partial E}|_{E=0}=\frac{1}{N}\sum_{i=1}^{N} 2n\mu\frac{\partial R_{i}(E,T)}{\partial E}|_{E=0}.
\end{equation}

The deviations $\frac{\displaystyle\partial R_{i}}{\displaystyle\partial T}$ and $\frac{\displaystyle\partial R_{i}}{\displaystyle\partial E}|_{E=0}$ can be obtained by numerical differential calculation, then $p(T)$ and $\chi (T)$
are obtained numerically. By changing the values of $J_{ab}$, $\Omega$ and $\sigma$, the effects of the interfacial coupling and the transverse field on the pyroelectric coefficient and dielectric susceptibility of the bilayer structure are investigated under the long-range interactions.\\

{\bf{III. RESULTS AND DISCUSSIONS}}\\

In order to make our computational simulation close to a real case, we hope we can include all the neighbors to our calculations. In Fig. 2, we plot the polarization of the ferroelectric bilayer as a function of the temperature at different cut-off approximation. The parameters, selected in Fig. 2, are $L_{a}=L_{b}=3$, $\Omega_{a}/J_{b}=\Omega_{b}/J_{b}=1.0$, $\sigma=3.0$, $J_{a}/J_{b}=0.5$, $J_{ab}/J_{b}=2.0$. $J_{b}$ is taken as the unit energy and $J_{a}$ is fixed as $0.5J_{b}$ in the whole calculation. From Fig. 2, we find that all the curves of different cut-off approximations trend to a limit. Therefore, we believe that the eighth-neighbor cut-off approximation will make our calculation reliable and practicable.\\

The effects of the strength of the interfacial coupling on the pyroelectric coefficient and dielectric susceptibility of the bilayer$(L_{a}=L_{b}=8)$ are given in Fig. 3(a) and Fig. 3(b), respectively. The transverse field and the parameter of the long-range interaction are fixed as $\Omega_{a}/J_{b}=\Omega_{b}/J_{b}=1.0$ and $\sigma=3.0$ in Fig. 3, respectively. It is shown that the phase transition temperature of the bilayer increases with the increase of the magnitude of the long-range interfacial coupling. From Fig. 3(a), we can see that there exist two peaks of the pyroelectric coefficient with the increase of the temperature for weak interfacial coupling. The first is a round peak, while the second is a sharp peak. This is different from the pyroelectric coefficient of the ferroelectric superlattice, where two peaks are found only for large-period superlattice, in which the long-range effect is not investigated.[14] This show that long-range interaction should be considered in study dielectric properties of ferroelectric materials. For strong interfacial coupling, three peaks of the pyroelectric coefficient are found, which correspond to the phase transition of slab A, slab B and the interface respectively. For weak interfacial coupling, the peak of the pyroelectric coefficient of the interface is merged into the peaks of the pyroelectric coefficient of slab A and slab B. In Fig. 3(b), there also exist two and three peaks of the susceptibility for weak and strong interfacial interaction, respectively. We believe that the reason is the same as the case of the pyroelectric coefficient.\\

The sharp peak mostly occurs at a higher temperature (the last peak from the low temperature to high temperature). This phenomenon is reasonable because the peak indicates the transition point of the whole thin film. The broad and round peak occurs at a low temperature (not the last peak). When one layer transfers to be disordered, the other layer, which is still in the ordered state, will affect it due to long-range interaction effect.\\

The effects of the transverse field on the pyroelectric coefficient and the susceptibility of the bilayer are shown in Fig. 4. The parameters in Fig. 4 are selected as: $L_{a}=L_{b}=8$, $J_{ab}/J_{b}=4.0$, $\sigma=3.0$. From Fig. 4, we can conclude that the phase transition temperature decreases with the increase of the transverse field, which is similar to the case of ferroelectric interfacial coupling in superlattice. The peaks of the pyroelectric coefficient and the susceptibility at low temperature will disappear at a large transverse field. Slab $A$ will be disordered and slab $B$ remains ferroelectric at zero temperature as the increase of the transverse field. Therefore, the system will behavior as a single ferroelectric slab when the transverse field is large enough, and the peaks of the pyroelectric coefficient and the susceptibility at low temperature, which corresponding to the phase transition of slab $A$, will disappear naturally. These results show that the quantum effect cannot be neglected when the long-range interaction is taken into account.\\
   
We present the effects of the long-range interaction on the pyroelectric coefficient and the susceptibility of the bilayer for $L_{a}=L_{b}=8$ in Fig. 5(a) and Fig. 5(b), respectively. The interfacial coupling and the transverse field are chosen as $J_{ab}/J_{b}=4.0$, $\Omega_{a}/J_{b}=\Omega_{b}/J_{b}=1.0$. Three peaks of the pyroelectric coefficient and the susceptibility appear, which correspond to the phase transition of slab $A$, slab $B$ and the interface respectively are found. From Fig. 5, we find that the position of the last peak for smaller $\sigma$ is on the right side of the position of that for larger $\sigma$, which indicates that the phase transition temperature of the bilayer increase with the increase of the long-range interaction. The pyroelectric coefficient and the susceptibility decrease with the increase of the magnitude of the long-range interaction before the phase transition. For thin ferroelectric bilayer($L_{a}=L_{b}=3$) in Fig. 6, the long-range interaction will lead to the disappearance of some peaks of the pyroelectric coefficient and susceptibility, and there is only one sharp pyroelectric and susceptibility peak for strong long-range interactions($\sigma=3.0$). The results above show that, when the film coupled with the larger long-range interaction (smaller $\sigma$), the size-effect become apparent under certain conditions ($J_{ab}=4.0, \Omega=1.0$). When the film become thinner and thinner, the pyroelectric coefficient and dielectric susceptibility peaks tends to be smoothen down. Until the film thickness decreases to a certain values, the two peaks disappear. We believe that the long-range interaction couple makes Slab A, Slab B and the interface of the thinner film combined into a whole body, which will contribute to a single effect. But when the film is considered with the weak long-range interaction(larger $\sigma$, or $\sigma\longrightarrow\infty $), the size-effect is not so obvious. This may be the reason why, in some certain experiments, the two or three peaks do not appear.\\

The phase transition temperatures as a function of the thickness $m$ ($L_{a}=L_{b}=m$) of the bilayer are shown in Fig. 7 for different values of $\sigma$. With the increase of the thickness $m$, the phase transition temperature will first increase, then keep invariable. At the same thickness of the bilayer, the long-range interaction will heighten the phase transition temperature, which is similar to the results of ferroelectric film.\\
 
In summary, we investigate the effects of the interfacial coupling and the transverse field on the pyroelectric and dielectric properties of the ferroelectric bilayer structure with the long-range interaction. The calculated results show that the interfacial coupling plays an important role on a ferroelectric bilayer structure. We obtain that: (1) When the long-range interaction is taken into consideration, with the increase of the magnitude of the interfacial coupling and the decrease of the transverse field, the phase transition temperature increases. (2) There exist three peaks of the pyroelectric coefficient and susceptibility for strong interfacial interaction, weak long-range interactions and low transverse field. With the increase of temperature for weak interfacial coupling one of the three peaks of the pyroelectric coefficient and the susceptibility will be merged into the other two peaks. (3) Under strong long-range interaction and a given transverse field and interfacial coupling, we find an obvious size effect on the pyroelectric coefficient and susceptibility. The pyroelectric coefficient and susceptibility peaks are more smooth and lower than those of the thicker film. (4) The phase transition temperature will increase with the thickness of the bilayer firstly, then keep almost invariable with the increase of the thickness, and the long-range interaction will cause the increase of the phase transition temperature.\\

{\bf {ACKNOWLEDGMENTS:}}

This work was supported by the National Natural Science Foundation of China under the Grant No.10174049 and the Natural Science Foundation of Jiangsu Education Committee of China.

\newpage
\begin{center}{CAPTION OF FIGURES}\end{center}
Fig. 1:\\
The schematic representation of a ferroelectric bilayer composed of two different ferroelectric slabs with a ferroelectric interfacial coupling.\\
Fig. 2:\\
The polarization of the bilayer as a function of temperature for different cut-off approximations.\\
Fig. 3\\
The pyroelectric coefficient and susceptibility of the bilayer as a function of temperature for different strength of the interfacial couplings.\\
Fig. 4\\
The pyroelectric coefficient and susceptibility of the bilayer as a function of temperature for different transverse fields.\\
Fig. 5\\
The pyroelectric coefficient and susceptibility of the bilayer ($L_{a}=L_{b}=8$) as a function of temperature for different values of the parameter of the long-range interactions.\\
Fig. 6\\
The pyroelectric coefficient and susceptibility of the thin bilayer ($L_{a}=L_{b}=3$) as a function of temperature for different values of the parameter of the long-range interactions.\\
Fig. 7\\
The plots of the phase transition temperature versus the thickness $m$ of the $A_{m}B_{m}$ bilayer structure for selected values of the parameter of the long-range interaction.

\newpage
\vfil\includegraphics[scale=0.7]{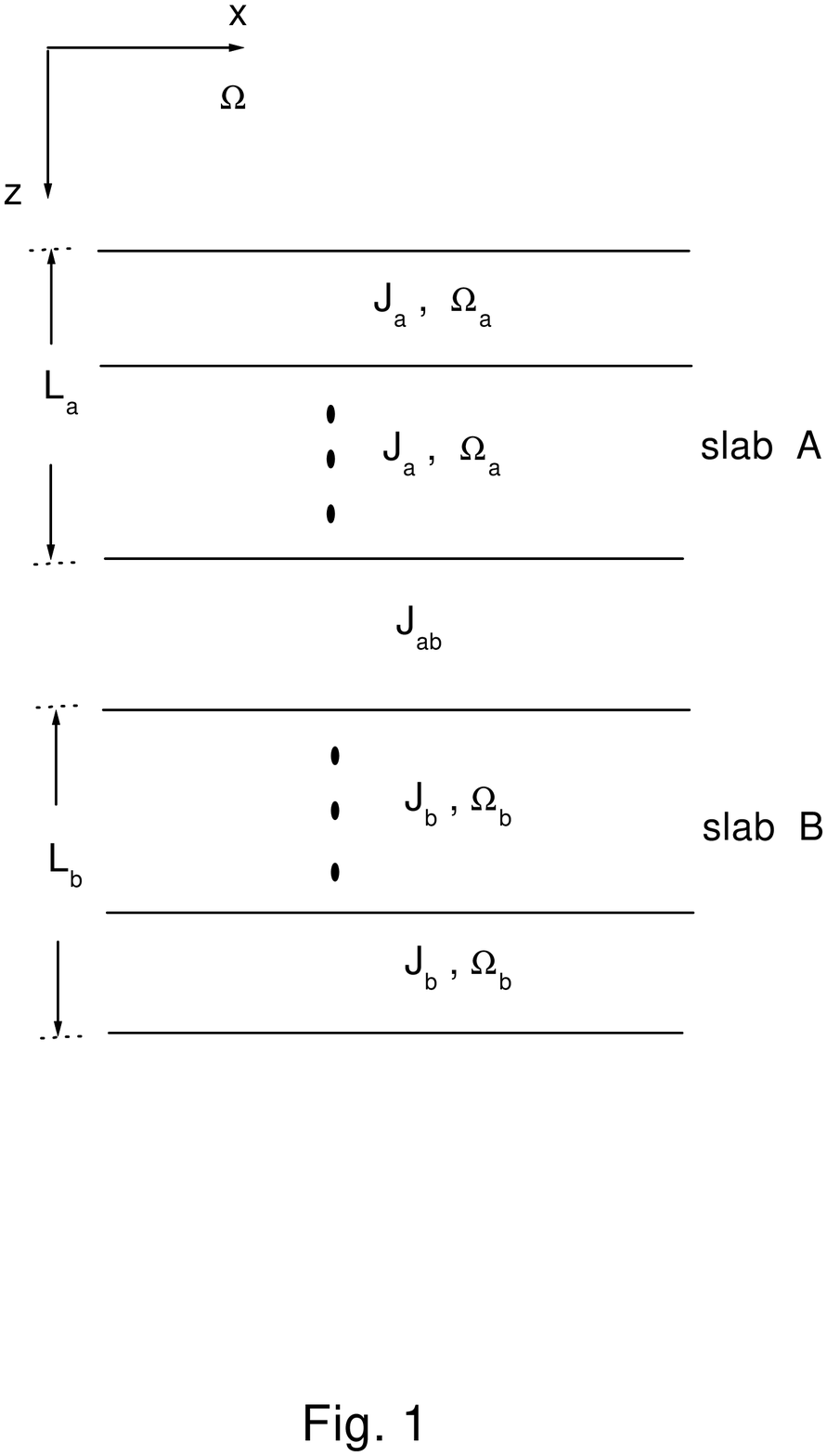}\vfil

\newpage
\vfil\includegraphics[scale=0.7]{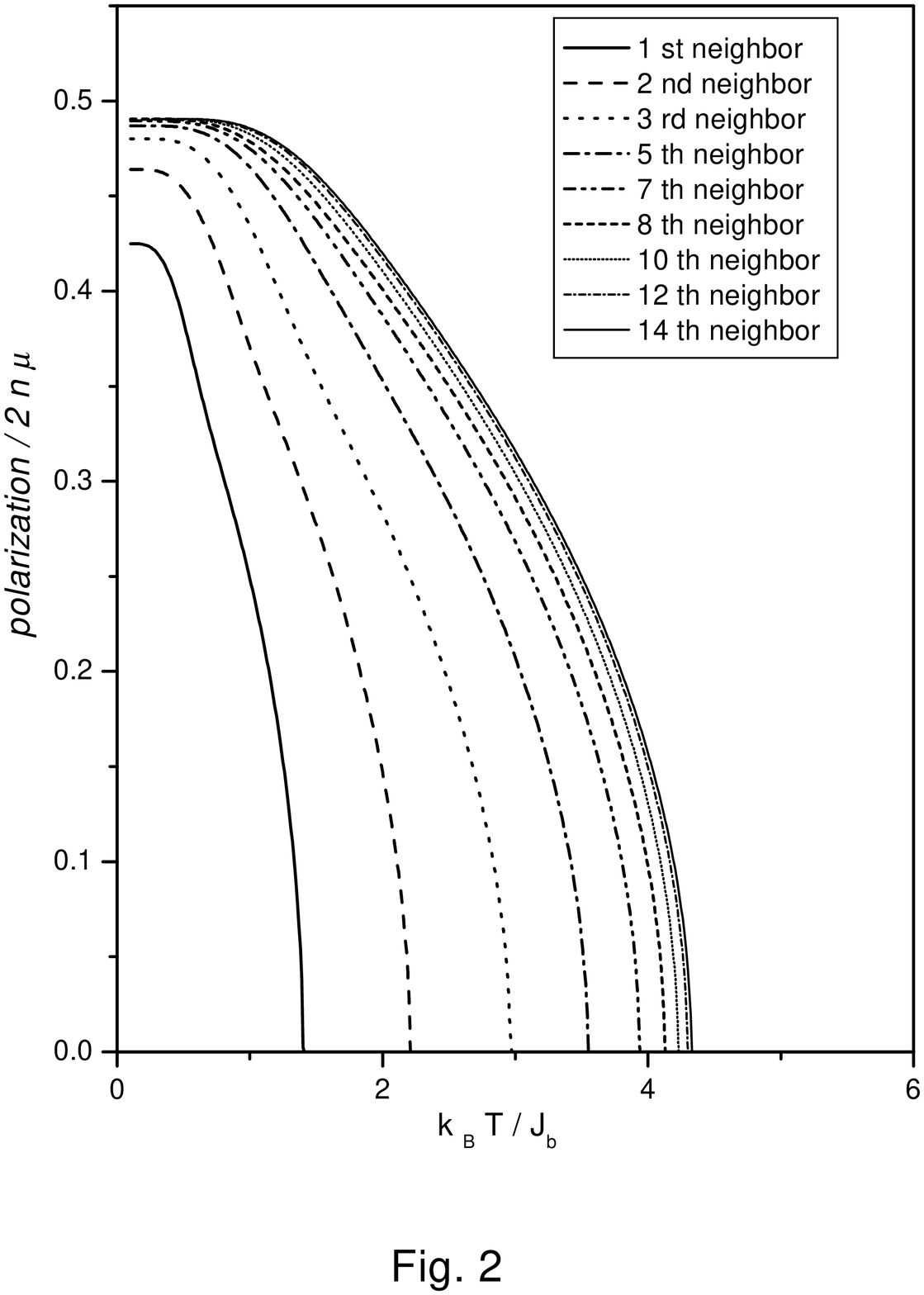}\vfil

\newpage
\vfil\includegraphics[scale=0.7]{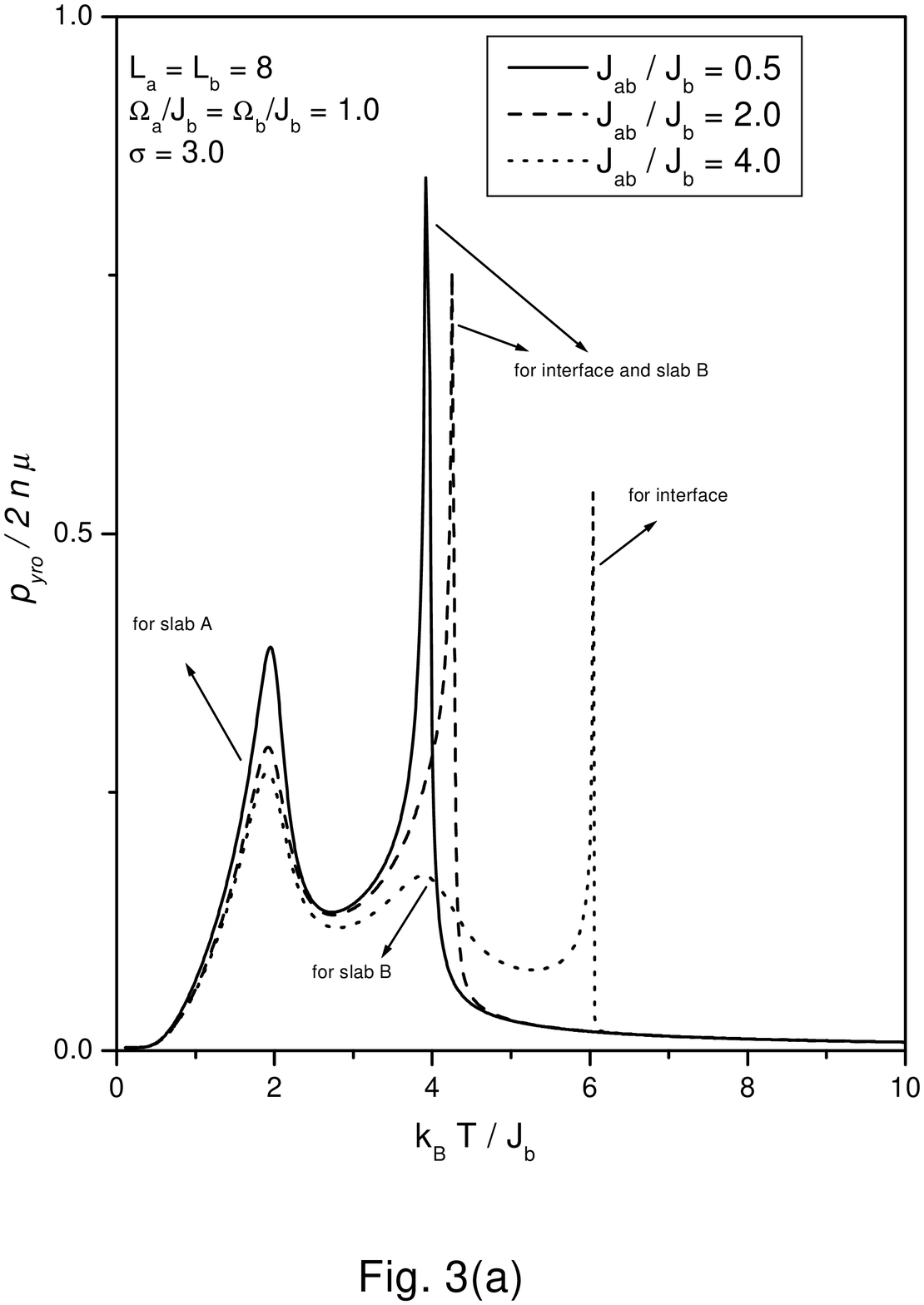}\vfil

\newpage
\vfil\includegraphics[scale=0.7]{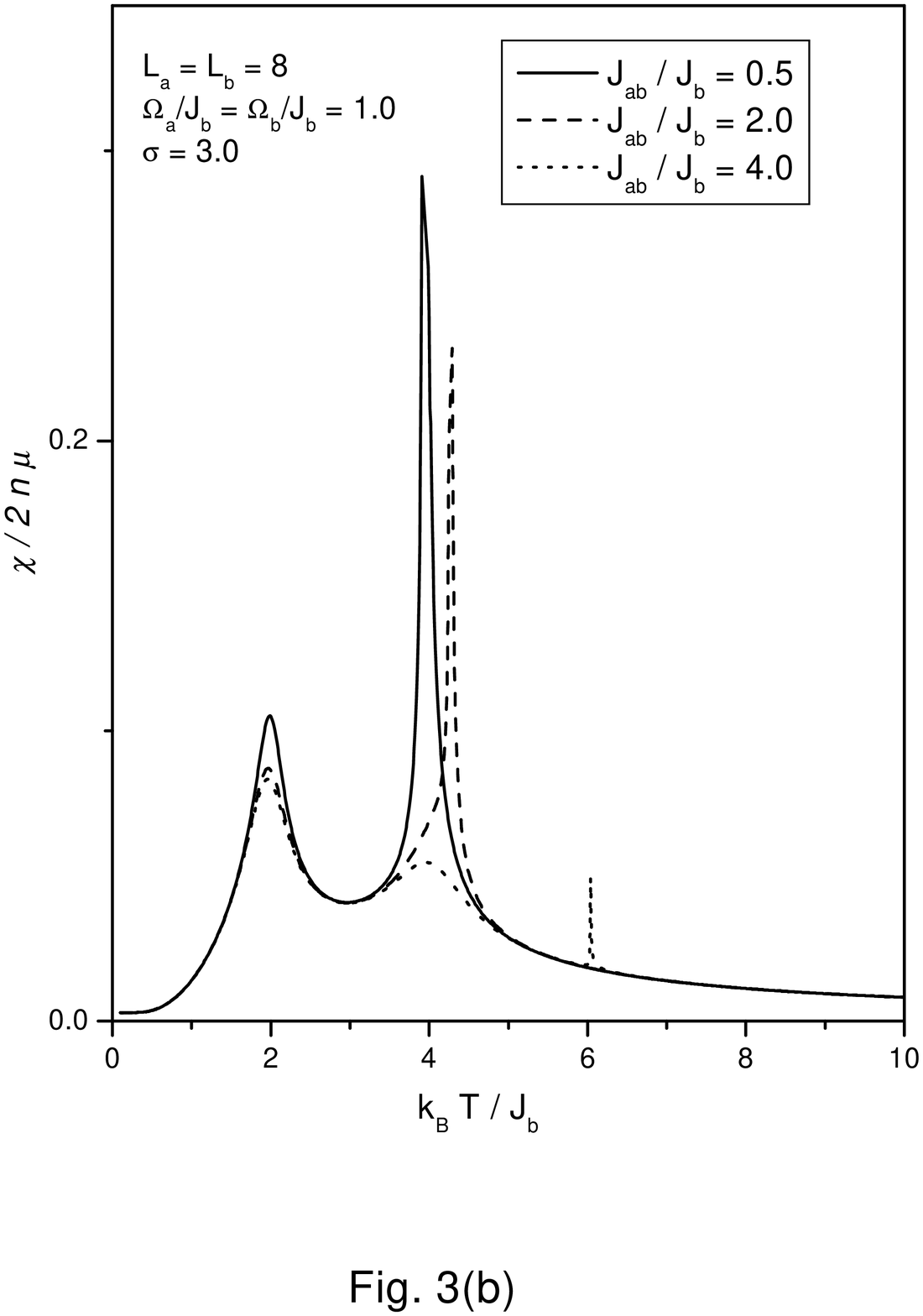}\vfil

\newpage
\vfil\includegraphics[scale=0.7]{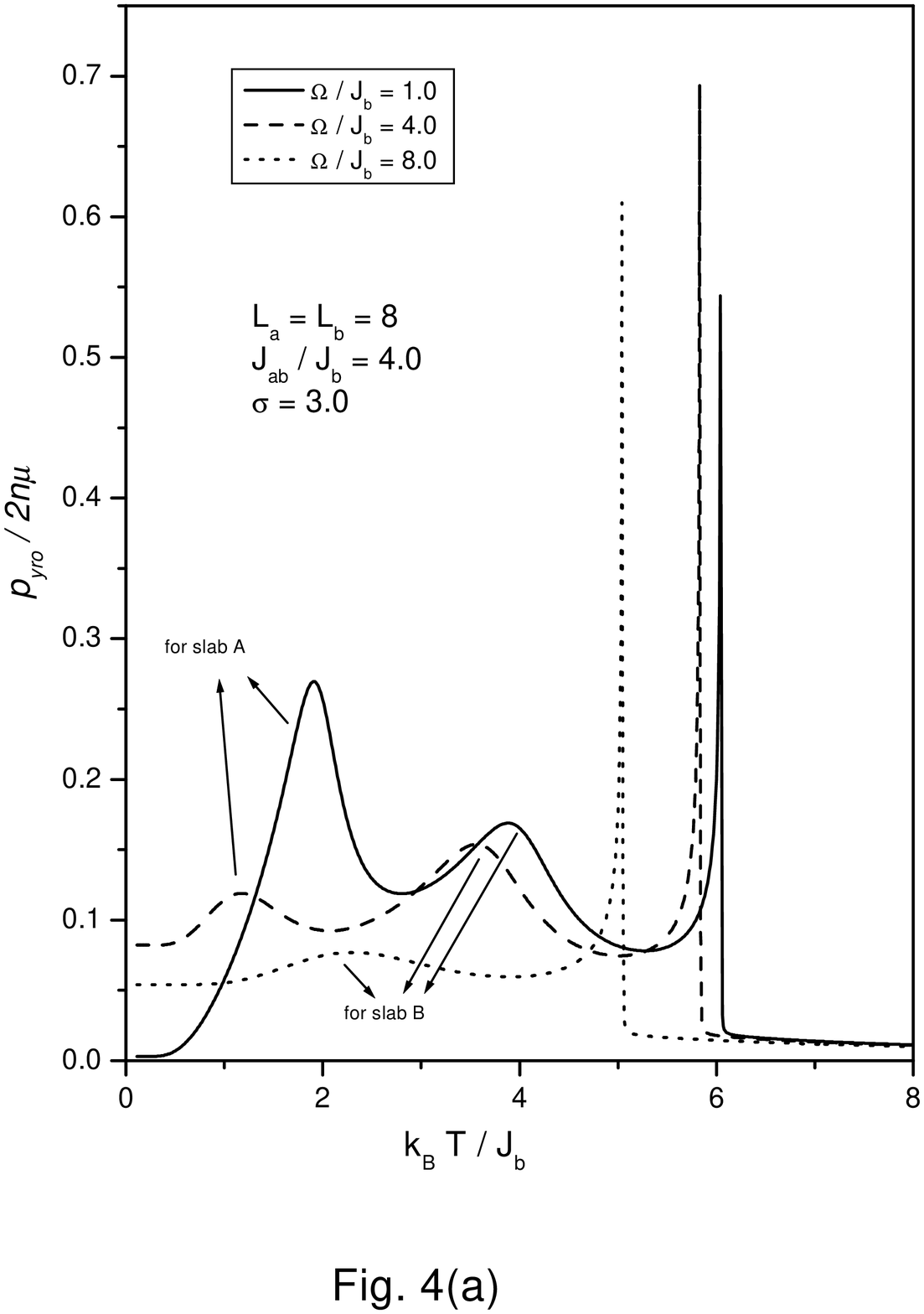}\vfil

\newpage
\vfil\includegraphics[scale=0.7]{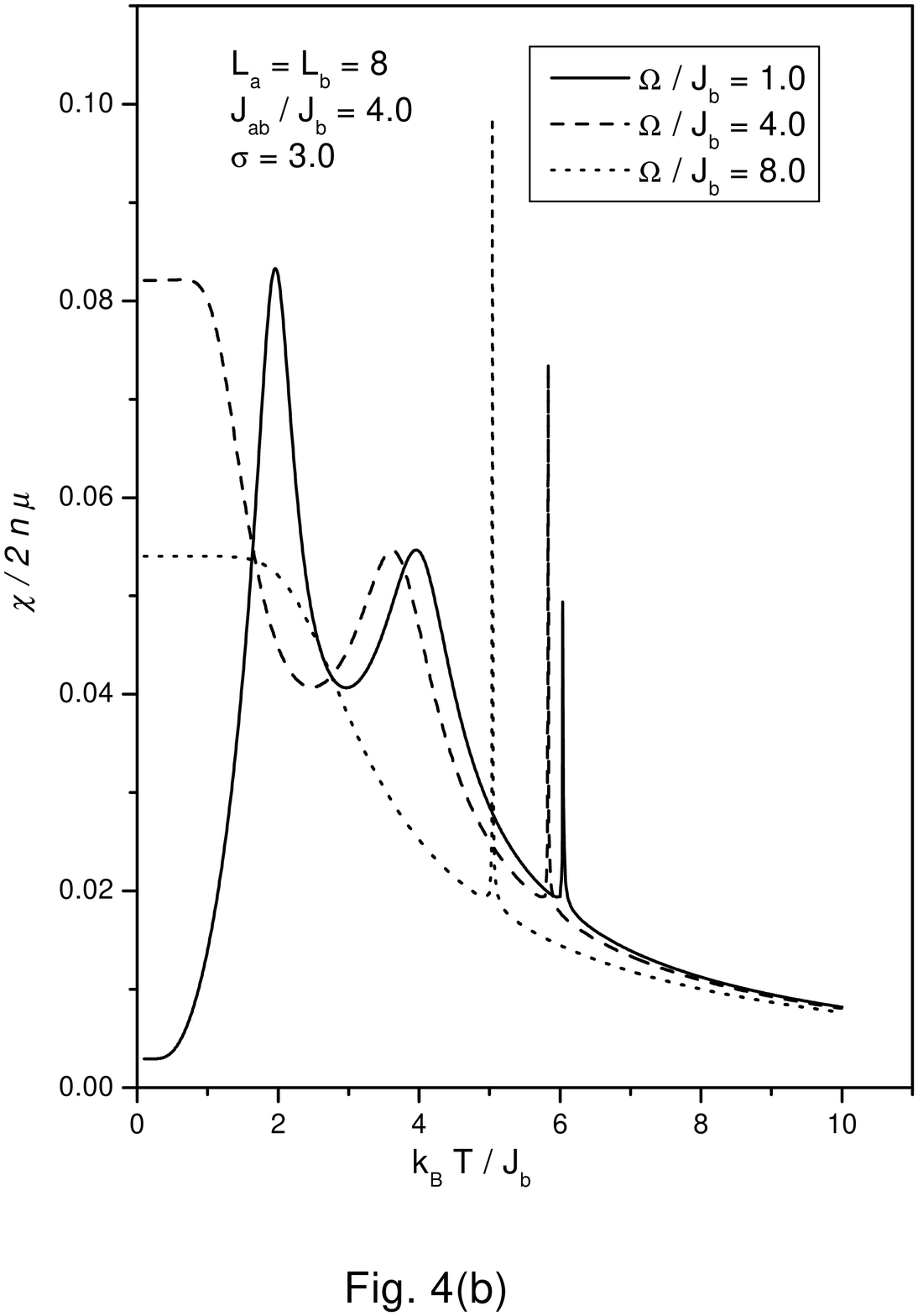}\vfil

\newpage
\vfil\includegraphics[scale=0.7]{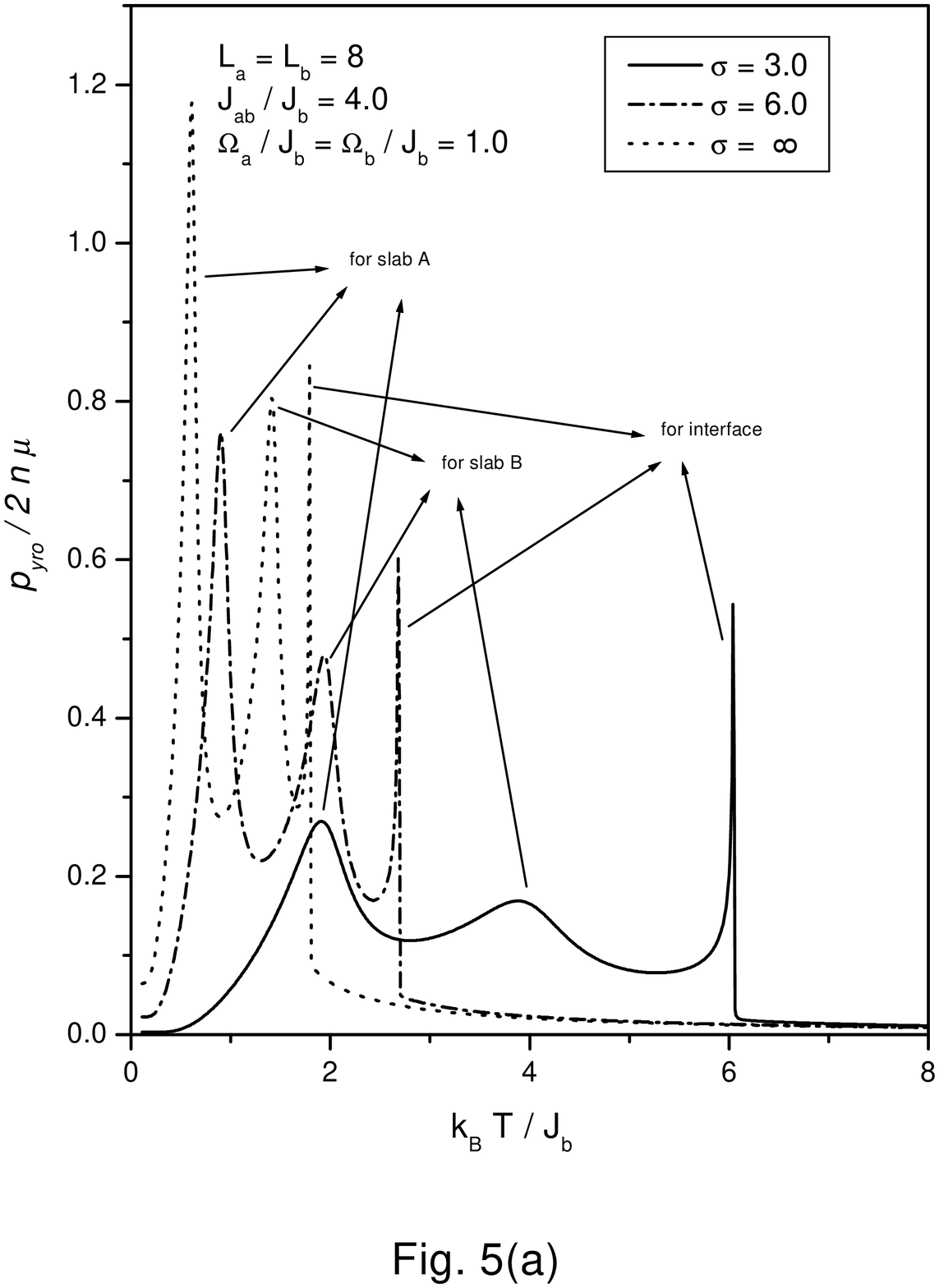}\vfil

\newpage
\vfil\includegraphics[scale=0.7]{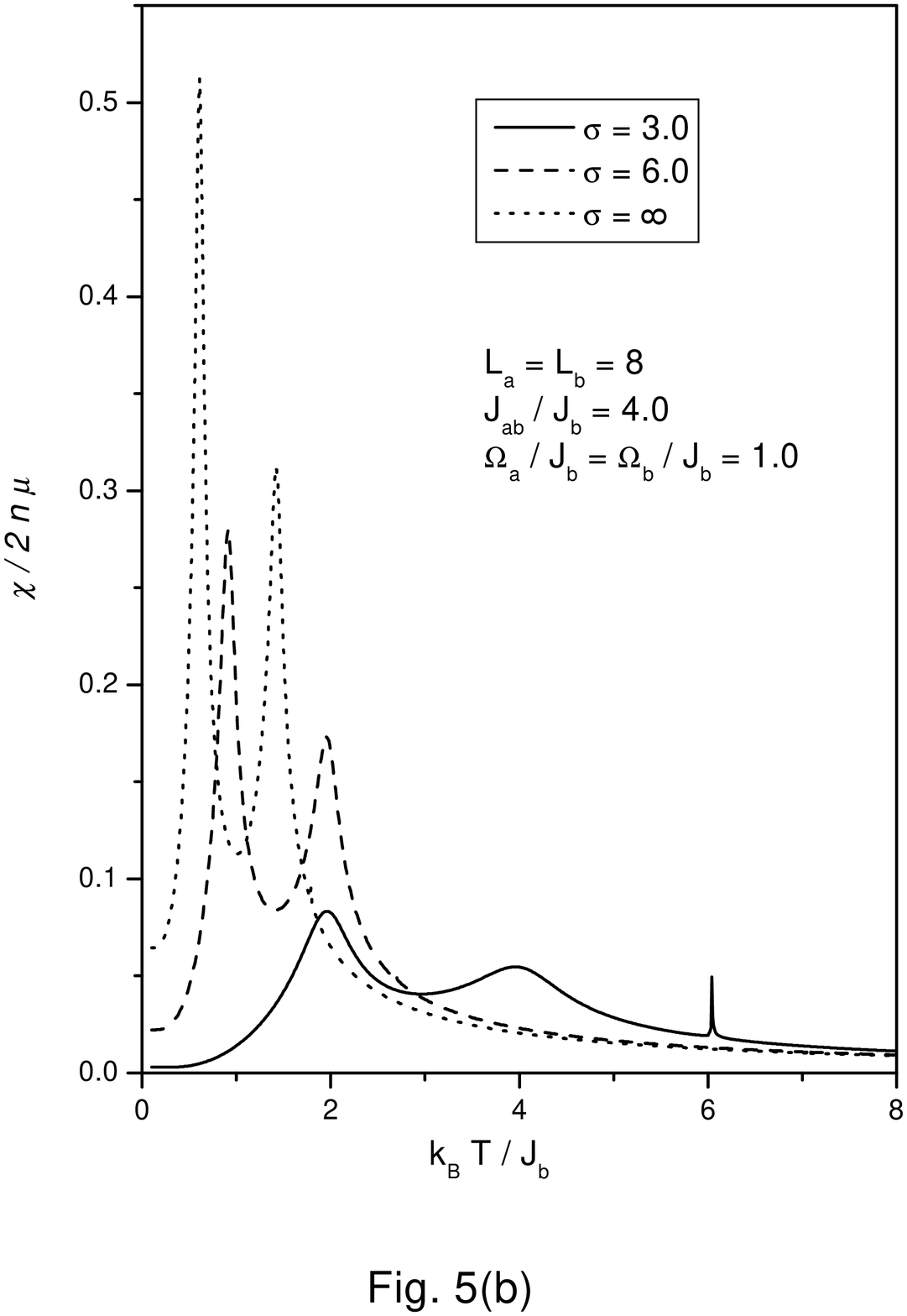}\vfil

\newpage
\vfil\includegraphics[scale=0.7]{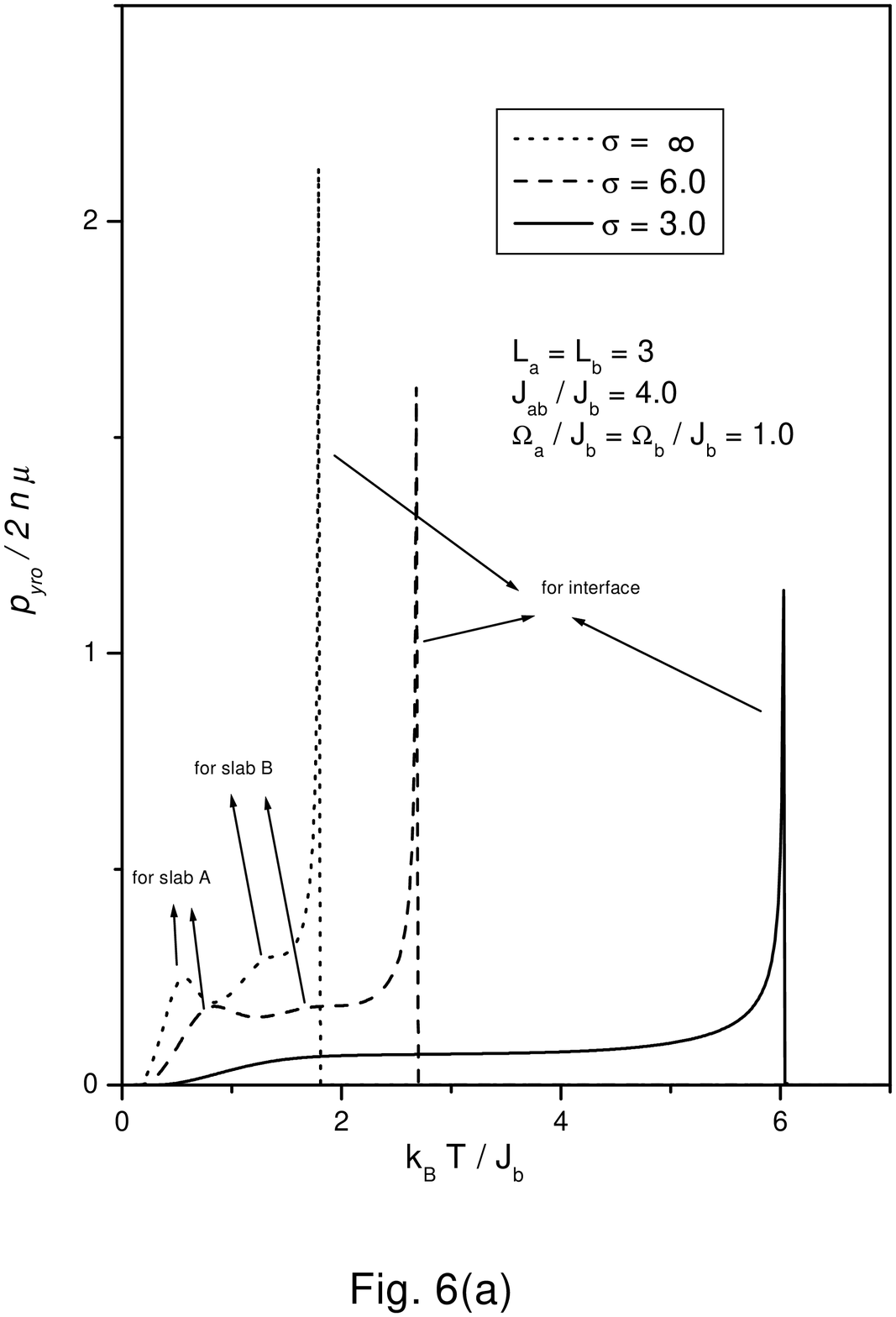}\vfil

\newpage
\vfil\includegraphics[scale=0.7]{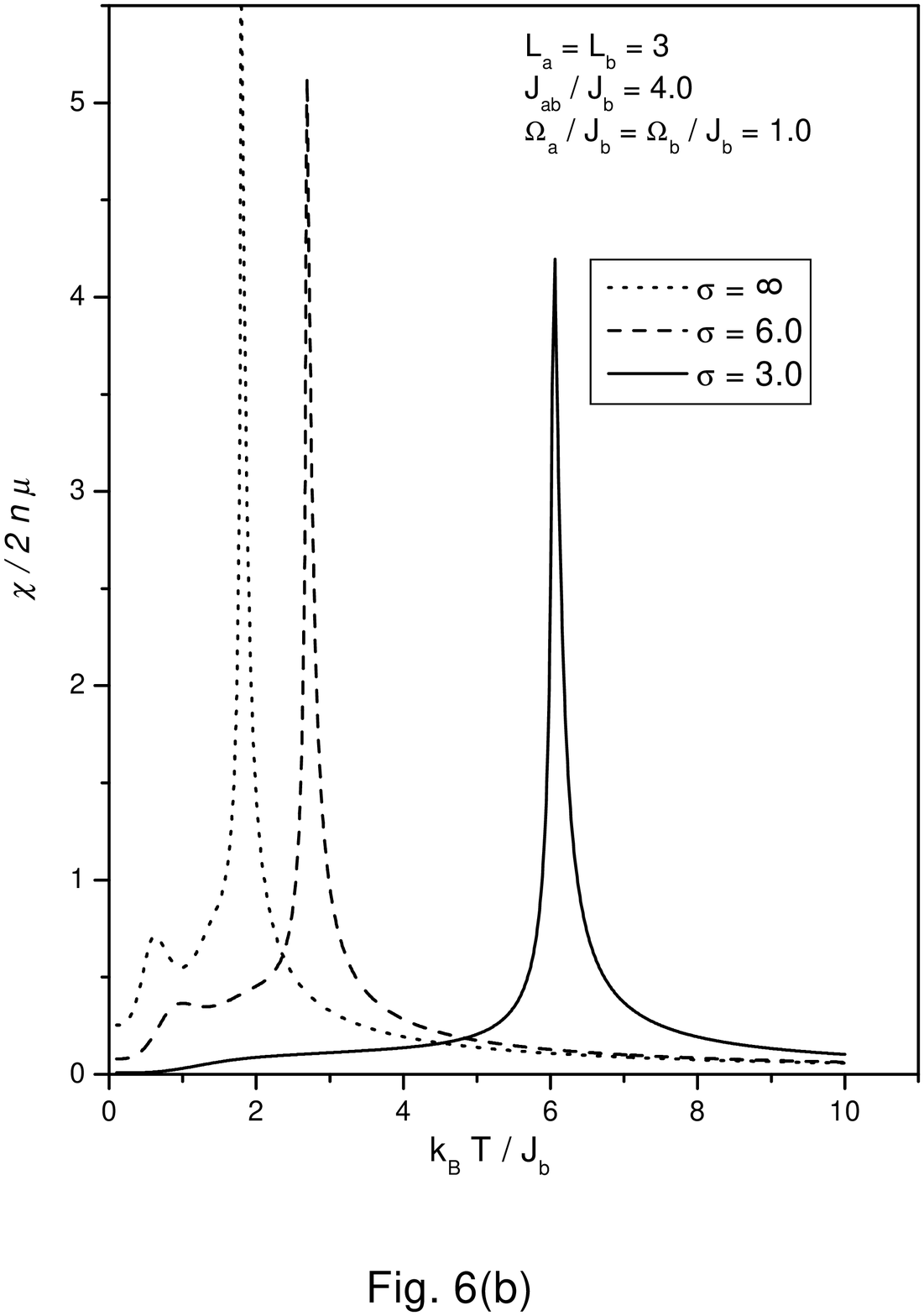}\vfil

\newpage
\vfil\includegraphics[scale=0.7]{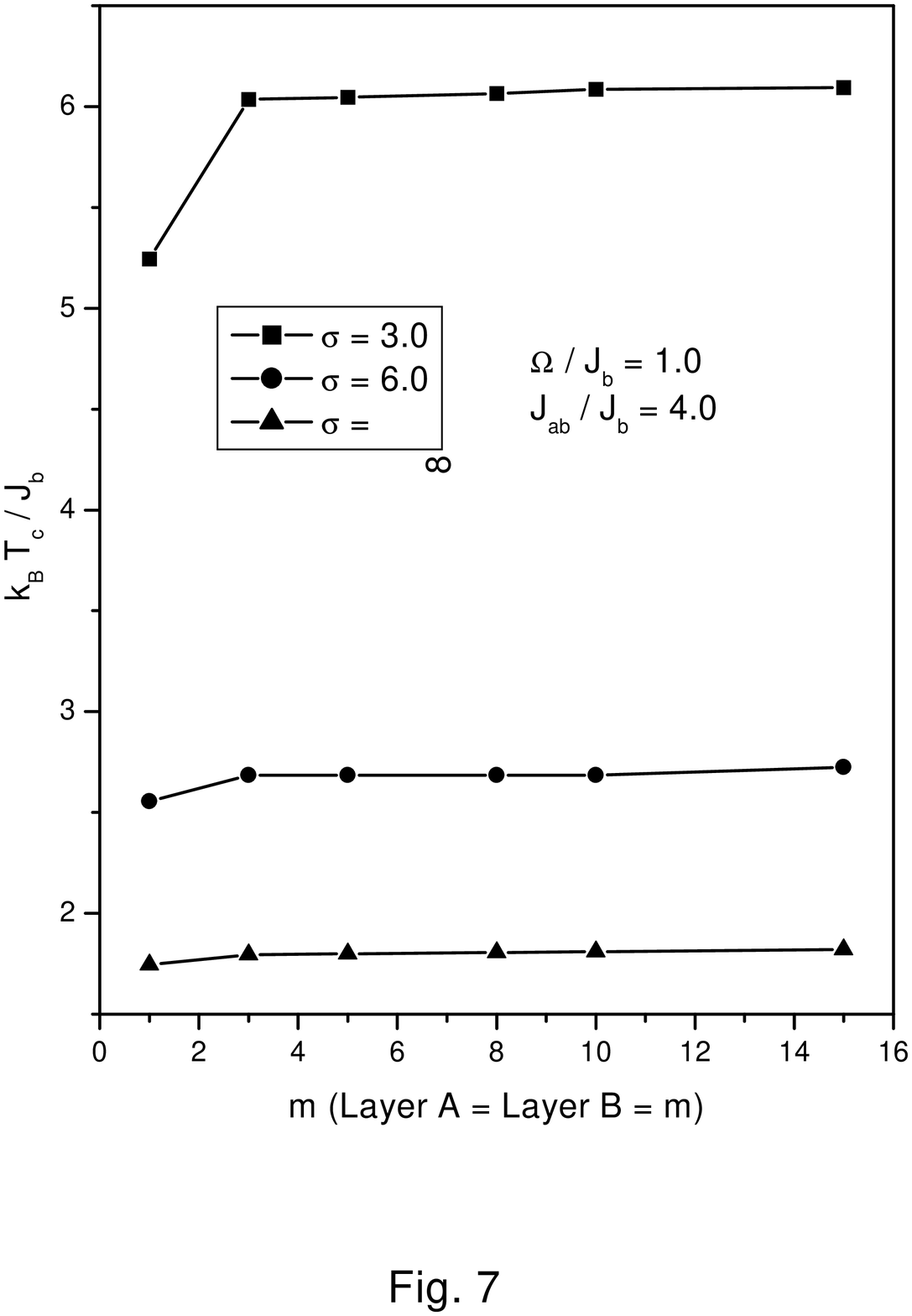}\vfil

\end{document}